\documentstyle[12pt]{article}
\def\lapp{{\ \lower 0.6ex \hbox{$\buildrel<\over\sim$}\ }}
\def\gapp{{\ \lower 0.6ex \hbox{$\buildrel>\over\sim$}\ }}
\newcommand{\gapproxeq}{\lower .7ex\hbox{$\;\stackrel{\textstyle >}{\sim}\;$}}
\newcommand{\lapproxeq}{\lower .7ex\hbox{$\;\stackrel{\textstyle <}{\sim}\;$}}
\def\to{\rightarrow}
\def\as{\alpha_s}

\def\pp{{\rm p\bar{p}}}

\def\tbar{{\bar{t}}}
\def\tt{t\bar{t}}

\def\GeV{{\rm GeV}}

\def\pb{{\rm pb}}

\def\ee{e^+e^-}
\def\gt{\Gamma_t}
\def\mt{m_t}
\def\cdfold{1}
\def\dzero{2}
\def\lep{3}
\def\mirkes{4}
\def\os{5}
\def\fayardlhc{6}
\def\fayardatlas{7}
\def\sumino{8}
\def\igoetal{9}
\def\igo{10}
\def\cdf{11}

\begin{document}

\begin{titlepage}
\vspace*{-1cm}
\begin{flushright}
 UR-1359 \\
ER-40685-809 \\
    May 1994 \\
\end{flushright}
\vskip 2.5cm
\begin{center}
{HOW TO MEASURE $m_t$:  A BRIEF OVERVIEW}\footnote{
presented at ``Electroweak Interactions and Unified Theories,'' XXIXth
Rencontres de Moriond, M\'eribel, Savoie, France, March 12-19, 1994;
proceedings to be published by Editions Fronti\`eres.}
\vskip 1.cm
{Lynne H. Orr}\footnote{
Work supported in part by the U.S. DOE, grant DE-FG02-91ER40685.}\\
\vskip .1cm
{\it Department of Physics Astronomy\\
University of Rochester\\
Rochester, NY 14627, USA } \\
\vskip .3cm
\vskip 1cm
\end{center}
\vskip 3 cm
\begin{flushleft}
ABSTRACT\\

\end{flushleft}
\noindent
This talk is an overview of prospects for measuring the top mass $m_t$
at present and future colliders.
Methods for extracting $m_t$ appropriate to experiments at the
Tevatron, Large Hadron Collider, and Next Linear Collider are discussed,
with examples given for each.
Sources of systematic uncertainties specific to each method and experiment are
identified.

\end{titlepage}

{\bf\noindent 1. Introduction}
\vglue 0.2cm
Discovery of the top quark will be followed immediately by
attempts to measure its mass $\mt$.
What we know so far is suggestive but not
conclusive.  As of this conference, direct searches have been
unsuccessful, leading to lower
limits on $\mt$ of $113\ \GeV$ and $131\ \GeV$ from  CDF$^{\cdfold)}$ and
D0$^{\dzero)}$, respectively.
Global fits to precision electroweak measurements at the $Z^0$ from
LEP and SLC imply$^{\lep)}$ $\mt=174^{+11+17}_{-12-19}\ \GeV$.
The top mass is of interest
not only as a parameter of the Standard Model (SM).
Since top is heavier by
far than the other quarks, it may be uniquely able to shed light on the origin
of mass, and many quantities both within
and beyond the SM depend on $\mt$, some quite sensitively.  Hence even our
ability to make predictions to test extensions or alternatives to the SM
depend on knowing the top mass.

Measuring the mass of the top quark is not so straightforward as for
the other quarks.
Because top is so heavy, it has a large width ($\gt\approx1\ \GeV$ for
$\mt=150\ \GeV$; $\gt$ increases as $\mt^3$) and it decays before
it can form hadrons.  The absence of a sharp $\tt$ resonance
makes $\mt$ hard to measure.
In this talk we discuss methods for measuring $\mt$ that
we do have at our disposal at present and future
hadron and $\ee$ colliders.  This is not meant to be a comprehensive review;
rather we present a brief overview with examples of
available methods.

Methods for measuring $\mt$ fall into three categories:
({\it i.})
measuring the production cross section and comparing
observed event rates with those expected;
({\it ii.})
reconstructing the top quark's four-momentum from the momenta
of its decay products; and
({\it iii.)}
measuring other distributions that are sensitive to $\mt$.
We will see examples of each in the discussions below.  Exactly how well
we can do will depend, among other things, on the value of $\mt$ itself.
The heavier top
is, the harder it is to produce, and the larger is its intrinsic width.

Before discussing specific experiments,
some comments about top detection modes are in order.
The $t$ quark decays
to a real $W$ boson and a $b$ quark nearly 100\% of the time.
Since top is predominatly produced in pairs, its
detection modes are characterized by the $W$ decays.
\vspace{-.2\baselineskip}
\begin{itemize}
\item
In the {\it dilepton} mode, each of the $W$'s decays to an $e$ or a $\mu$,
and such events contain two isolated charged leptons, two $b$ jets, and
missing energy from two undetected neutrinos.  This is the cleanest
channel, with the lowest backgrounds.  Unfortunately it has a
low branching ratio ($\sim 5\%$), and the neutrinos
prohibit top momentum reconstruction.
\vspace{-.2\baselineskip}
\item
In the {\it all jets} mode, both $W$'s decay
hadronically.  These events contain six jets, and so all of the decay products
from both top quarks are, in principle, detectable.  With a branching ratio of
45\%, this mode occurs at the largest rate.  However, at
hadron colliders, large QCD backgrounds can overwhelm the signal.
\vspace{-.2\baselineskip}
\item
A compromise is struck in the {\it lepton+jets} mode, with hadronic and
one leptonic $W$ decay.  The branching ratio is appreciable
(15\%) and the events are characterized by one isolated
lepton, four jets (two from $b$'s), and missing energy from a
single neutrino.
One of the top
momenta can be fully reconstructed from jets.
\end{itemize}
\vspace{-.2\baselineskip}
We will see below how the detection channels and methods of measurement are
best combined
in particular experiments, and what this implies about sources of
systematic uncertainties.

\vglue 0.3cm
{\bf\noindent 2. Top mass measurement at the Tevatron}
\vglue 0.2cm

For a top quark mass on the order of $150\ \GeV$,
the Tevatron $\pp$ collider will produce
about $10^4$ $\tt$ pairs in $10^3\ \pb^{-1}$ of integrated
luminosity.  The challenge will be to make the most of the relatively
modest numbers of events left over after branching ratios and cuts to reduce
background are taken into account.
Of the mass determination methods mentioned above, reconstructing the
top momentum has the best prospects here.  Measuring the production
cross section is useful for setting lower limits on $\mt$ in the
absence of a significant signal, but theoretical uncertainties are too
large to make it of much use for measuring $\mt$
once a signal has been obtained.  And the event rate will be insufficient for
measuring distributions other than that of the reconstructed mass.

Reconstructing the top momentum at the Tevatron will be best achieved in the
lepton+jets channel.  Requiring a single lepton cuts down on QCD backgrounds
while allowing full reconstruction of one of the top momenta.
Furthermore, the other top momentum
can be reconstructed (with a two-fold ambiguity)
by attributing missing transverse energy to the undetected neutrino.
This method is discussed, {\it e.g.}, in Ref.\ [\mirkes], from which we show
some
results for $\mt=150\ \GeV$  in Figure 1.  Kinematic cuts and energy smearing
have been applied, but no hadronization or detector effects are included.
At least one $b$ tag is required.  Figs.\ 1(a) and 1(b) show, respectively,
the reconstructed $bl\nu$ mass (using the tagged $b$) and the
reconstructed 3-jet mass (with a $W$ mass constraint on two of the jets).
Combinatorial backgrounds from wrong jet combinations are included in the
$\tt$ curves and the $W$+jets background is shown as the dashed line.
A narrower signal peak can be obtained by combining the two methods,
as shown in Fig.\ 1(c).  It
shows the distribution of the mean of the reconstructed--mass pair which
gives the closest two values of $\mt$, with their difference
constrained to be $<50\ \GeV$.
\begin{figure}
\vspace{7cm}
\caption{Reconstructed top mass at the Tevatron in lepton+jets channel.
$tt$ signal (solid line) and $W$+jets background (dashed line) shown
for (a) leptonically decaying top; (b) hadronically decaying top; (c) mean
of (a) and (b) with mass difference constraint.  From Ref.\ [\mirkes].}
\end{figure}

The sources of systematic uncertainties
include, from the theory, higher order QCD effects in predictions for both
signal and backgrounds, and on the experimental side, jet definition
and energy scale, detector efficiencies and acceptance, and effects
related to $b$-tagging.  In addition, extra gluons can
be radiated before and after top decays and can lead to further
ambiguities in mass reconstruction.$^{\os)}$
{}From such analyses we can expect eventually to obtain a top mass
measurement at the Tevatron to about 10--15 GeV, depending on $\mt$.

\vspace{0.3cm}
{\bf\noindent 3. Top mass measurement at the Large Hadron Collider}
\vglue 0.2cm

The LHC will be a top factory:  with a
yearly integrated luminosity of $10^4\ \pb^{-1}$, it will produce on the
order of
$10^7\ \tt$ pairs each year.  Unfortunately, with high luminosity
we get  multiple interactions, which leads to problems for any measurement
involving jets because of extra hadronic activity.
There are two ways to get around this for purposes of measuring
$\mt$:$^{\fayardlhc)}$  we can run at reduced luminosity, and we can
run at design luminsity, but avoid using jets.

\vspace{0.2cm}
{\it\noindent 3.1 Mass reconstruction at low luminosity}
\vglue 0.2cm
The cross section for top production is large enough at LHC energies that
we can afford to reduce the luminosity by a factor of ten to ameliorate some
of the problems associated with high luminosity, and still have an
appreciable event rate.
We can then use the lepton+jets channel to reconstruct the top
momentum as at the Tevatron.  Figure 2 shows the three-jet invariant
mass distribution obtained in Ref.\ [\fayardlhc] for top events at the LHC,
for $\mt=130\ \GeV$ and $\mt=200\ \GeV$.  The distributions show
clear peaks at the top mass.   In addition to kinematic cuts and a $W$ mass
constraint, $b$ fragmentation and detector effects are included.
Backgrounds can be reduced further by using $b$ tagging, though some signal
events are lost in the process as well.
This measurement will have larger systematic than statistical uncertainties
(the large fluctuations shown are the result of limited Monte Carlo
statistics),
and the dominant sources of systematic uncertainties will be those associated
with measuring jet energies and possibly $b$ tagging.
These include initial state radiation and hadronic activity
from the underlying event, the jet cone algorithm, and especially $b$
fragmentation and calorimeter response.
\begin{figure}
\vspace{7cm}
\caption{Reconstructed top mass $m_{jjj}$ at reduced-luminosity LHC using the
lepton+jets channel.  (a) $\mt=130\ \GeV$.  (b) $\mt=200\ \GeV$.
{}From Ref.\ [\fayardlhc].}
\end{figure}

\vspace{0.2cm}
{\it\noindent 3.2 Sequential $ll'$ mass distribution at design luminosity}
\vglue 0.2cm
At design luminosity at the LHC, dealing with jets in top events will be
difficult.
However, the $\tt$ event rate is so large
that we can hope to
find an easily measured distribution that
is sensitive to $\mt$ but does not get spoiled by multiple interactions.
Such a distribution can be obtained$^{\fayardlhc)}$ in the clean dilepton
channel if we
require that one of the $b$'s also decays leptonically,
{\it e.g.}, $t\to W^+(\to l^+\nu)b(\to cl^-\bar\nu)$ with
$\tbar\to W^-(\to l^-\bar\nu)b$.
The distribution of the invariant mass  of the
opposite-sign pair of leptons that come from the same top
({\it e.g.}, the $l^+$ and $l^-$ from the $t$ in the above example)
has the desired properties.
Hence one searches for two isolated leptons and one additional non-isolated
lepton from the $b$ decay.  In Figure 3 (from Ref.\ [\fayardlhc]),
which shows $m_{ll'}$ distributions for $\mt=130$ and $200\ \GeV$, we see that
the peak and mean values increase with $\mt$.  (Again, fluctuations are due to
limited Monte Carlo statistics.)
  Here the systematic effects
are dominated by the lepton energy scales and details of the $b$ fragmentation,
and there may be contributions from
dependence of $m_{ll'}$ on the transverse momentum of the top.
Ultimately, using such
distributions, we expect the sensitivity of top mass measurements that may be
achieved at the LHC to be a few GeV, and perhaps as low as
$2\ \GeV$.$^{\fayardatlas)}$
Note that the expected sensitivity here begins to be of the same order
of magnitude as the intrinsic width of the top quark.
\begin{figure}
\vspace{7cm}
\caption{Sequential $ll'$ mass distribution at LHC using the
dilepton channel with semileponic $b$ decay.  (a) $\mt=130\ \GeV$.
(b) $\mt=200\ \GeV$.  From Ref.\ [\fayardlhc].}
\end{figure}

\vspace{0.3cm}
{\bf\noindent 4. Top mass measurement at the Next Linear Collider}
\vglue 0.2cm

A precision measurement of $\mt$ requires the clean environment of a high
energy $\ee$ collider.  All three methods discussed in the introduction
can be used, and a measurement of the top mass to about a few hundred MeV
can be obtained.  We can study the top production cross section and top
momentum
distribution near
$\tt$ threshold and, at higher energies, reconstruct the top mass, to
obtain independent measurements of $\mt$.  (In practice, it is likely that
we will use the latter measurement to tell us at what energies to perform the
threshold studies.)

\vspace{0.2cm}
{\it\noindent 4.1 Production cross section at $\tt$ threshold}
\vglue 0.2cm
As mentioned above, the top width is too large for a narrow
toponium resonance peak to appear in the production cross section.
Nonetheless, there is some structure in the threshold region due to the
attractive Coulomb-like QCD potential between the $t$ and $\tbar$ quarks.
Near the $\tt$ threshold, the cross section as a function of
collision energy exhibits a bump which spreads out and eventually disappears
as $\mt$ increases.  The height and position of the bump behave similarly
when $\as$ {\it decreases} (see Figure 4(a), from Ref.\ [\sumino]); this is
because $\as$ determines the depth of the potential.  Thus  one obtains
correlated measurements of $\mt$ and $\as$.
Here  the all jets channel can be used to identify the top
signal.  Systematic uncertainties in this case arise from effects due to
initial state radiation, beam energy spread, and beamstrahlung, and are
related to understanding the energy of the hard collision.
\begin{figure}
\vspace{7cm}
\caption{Cross section and top momentum distribution near $\tt$ threshold at
NLC.  (a) Production cross section as a function of collision energy.  From
Ref.\ [\sumino].  (b) Top 3-momentum distribution for $\mt=149, 150, 151\
\GeV$.
{}From Ref.\ [\igoetal].}
\end{figure}

Another handle available near the $\tt$ threshold is the 3-momentum
distribution $d\sigma/d|\vec{p_t}|$
of the top quarks, which have some Fermi motion.
Here one can use the lepton+jets channel (to reduce combinatorial backgrounds)
and measure $d\sigma/d|\vec{p_t}|$ at fixed collision energy,
as shown in Fig.\ 4(b) from Ref.\ [\igoetal].
The magnitude of the 3-momentum  at which $d\sigma/d|\vec{p_t}|$
peaks (typically of order 10--20 GeV) is very sensitive to $\mt$ and can be
relatively insensitive to $\as$.  The systematic uncertainties here are related
to energy and momentum measurement and  arise
from hadronic effects and undetected sources of momentum imbalance.

Because of the insensitivity of the $|\vec{p_t}|$ distribution to $\as$,
the best prospects for measuring $\mt$ come from combining a threshold energy
scan with a measurement of $d\sigma/d|\vec{p_t}|$ at fixed energy near
threshold.  Such a combined measurement is expected to give $\mt$ to
a few hundred MeV, depending on the value of $\mt$.

\vspace{0.2cm}
{\it\noindent 4.2 Mass reconstruction at high energy}
\vglue 0.2cm
At energies well above $\tt$ threshold at the NLC, it will be straightforward
to reconstruct the top 4-momentum from its decay products.  Using either
the all jets or lepton+jets channel, three-jet invariant masses
can be reconstructed to obtain $\mt$.  Figure 5, from Ref.\ [\igo],
shows $m_{jjj}$ distributions for $\mt=150\ \GeV$ and
$\sqrt{s}=500\ \GeV$.  We see that the lepton+jets
channel gives a narrower distribution, but the all jets channel has a
larger overall rate.  Systematic effects here arise from  measurement of
missing energy, detector acceptances, jet resolution, and QCD-related effects,
such as ambiguities in assigning gluon jets in reconstructed
momenta.$^{\os)}$
\begin{figure}
\vspace{7cm}
\caption{
Reconstructed top mass at the NLC for $\mt=150\ \GeV$ and
$\protect\sqrt{s} =500\ \GeV$
in the lepton+jets (inner hist.) and all jets (outer
hist.) channels.
{}From Ref.\ [\igo].}
\end{figure}

Note that the systematic effects involved in the high energy mesurement are
completely different from those in the threshold measurement, so that
these methods give independent measurements of $\mt$.  (The threshold
measurement gives a smaller overall uncertainty.)

\vspace{0.3cm}
{\bf\noindent 5. Summary}
\vglue 0.2cm
We expect top mass measurements to improve markedly with each
new machine at which top is produced.  How well we can do depends on the
actual value of $\mt$, but the following numbers are reasonable estimates.
Mass reconstruction at the Tevatron
can be expected to give $\mt$ to about 10--15 GeV.  At the LHC, methods
such as mass reconstruction at low luminosity or measurement of the
sequential $l l'$ distribution at design luminosity should reduce uncertainties
in $\mt$ to a few, perhaps 2, GeV.  Finally, precision
measurements will be possible at the NLC via cross section
and top 3-momentum measurements at $\tt$ threshold, and mass reconstruction
at higher energies, resulting in an $\mt$ measurement to a few hundred
MeV.

{\it Note added:}\/ As of this writing (May 1994),
CDF has reported$^{\cdf)}$ seeing evidence for top quark production.
Assuming the excess
they see is indeed due to top, they obtain $\mt=174\pm10^{+13}_{-12}\ \GeV$
from fits to their data.
Their method for obtaining $\mt$ is similar in spirit to that
discussed in section 2 above; see Ref.\ [\cdf] for details.

\vspace{.5cm}
\noindent{REFERENCES}
\addtolength{\baselineskip}{-.3\baselineskip}

\noindent
\begin{itemize}
\item[{[\cdfold]}]
P.Tipton for the CDF collaboration, Proc.\ XVI International
Symposium on  Lepton-Photon
Interactions, Ithaca, NY, August 1993, to be published.
\vspace{-.4\baselineskip}
\item[{[\dzero]}]
S.\ Abachi {\it et al.,} Phys.\ Rev.\ Lett.\ {\bf 72}, 2138 (1994) and
J.\ Cochran, these proceedings.
\vspace{-.4\baselineskip}
\item[{[\lep]}]
B. Pietrzyk, these proceedings.
\vspace{-.4\baselineskip}
\item[{[\mirkes]}]
V.\ Barger, E.\ Mirkes, J.\ Ohnemus, and R.J.N.\ Phillips, MAD/PH/807,
Dec.\ 1993.
\vspace{-.4\baselineskip}
\item[{[\os]}]
L.H.\ Orr and W.J.\ Stirling, in preparation.
\vspace{-.4\baselineskip}
\item[{[\fayardlhc]}]
G.\ Unal and L.\ Fayard, in Proc.\ LHC Workshop, Aachen, ed.\ C. Jarlskog and
D.\ Rein, CERN-90-10 (1990), Vol.\ II, p.\ 360.
\vspace{-.4\baselineskip}
\item[{[\fayardatlas]}]
A.\ Mekki and L.\ Fayard, ATLAS note PHYS-NO-028, July 1993.
\vspace{-.4\baselineskip}
\item[{[\sumino]}]
K.\ Fujii, T.\ Matsui, and Y. Sumino, KEK-PREPRINT-93-125, Oct.\ 1993.
\vspace{-.4\baselineskip}
\item[{[\igoetal]}]
P.\ Igo-Kemenes, M. Martinez, R. Miquel, and S. Orteu,
in {\it Workshop on Physics and Experiments with Linear
$e^+e^-$ Colliders\/}, Proc.\ 2nd International Workshop, Waikoloa,
 HI, ed. F.A. Harris, {\it et al.}, World
Scientific, Singapore (1993), p.\ 457.
\vspace{-.4\baselineskip}
\item[{[\igo]}]
P.\ Igo-Kemenes, {\it ibid.,}\/ p.\ 95.
\vspace{-.4\baselineskip}
\item[{[\cdf]}]
F.\ Abe {\it et al.,} Fermilab-Pub-94/097-E, April 1994.
\end{itemize}

\end{document}